\documentclass[9pt]{article}
\usepackage{bm, amsthm, amsfonts, amsopn,amsmath,graphicx,hyperref, amssymb, cleveref}
\usepackage{format}
\usepackage{commands}
\usepackage{booktabs}
\usepackage{array}
\usepackage{hyperref}
\usepackage{float}
\usepackage{graphicx}
\usepackage{xcolor}
\usepackage{url}
\usepackage[per-mode=symbol]{siunitx}

\usepackage{xcolor}
\usepackage{listings}
\definecolor{codegreen}{rgb}{0,0.6,0}
\definecolor{codegray}{rgb}{0.5,0.5,0.5}
\definecolor{codepurple}{rgb}{0.58,0,0.82}
\definecolor{backcolour}{rgb}{0.95,0.95,0.92}
\definecolor{LightGray}{gray}{0.9}
\lstdefinestyle{mystyle}{
    backgroundcolor=\color{backcolour},   
    commentstyle=\color{codegray}\itshape,
    keywordstyle=\color{codegreen}\bfseries,
    numberstyle=\tiny\color{codegray},
    stringstyle=\color{codepurple},
    basicstyle=\ttfamily\footnotesize,
    breakatwhitespace=false,         
    breaklines=true,                 
    captionpos=b,                    
    keepspaces=true,                 
    numbers=left,                    
    numbersep=5pt,                  
    showspaces=false,                
    showstringspaces=false,
    showtabs=false,                  
    tabsize=2,
    frame=lines,
    rulecolor=\color{black},
    xleftmargin=2em,
    framexleftmargin=1.5em
}
\allowdisplaybreaks
\usepackage{capt-of}
\usepackage{multirow}
\usepackage{cuted}
\usepackage[normalem]{ulem} %
\usepackage{enumitem} %

\title{Equivariant Deep Equilibrium Models for Imaging Inverse Problems}
\threeauthors
 {Alexander Mehta}
	{University of California, Berkeley\\   
    alexmehta@berkeley.edu}
 {Ruangrawee Kitichotkul and Vivek K Goyal}
	{    Boston University  \\
     \{rkitich, goyal\}@bu.edu}{Juli\'an Tachella}{    CNRS \& ENS Lyon \\ 
     julian.tachella@cnrs.fr}
\begin{document}
\maketitle
\begin{strip}
    \vskip -5em
    \centering
    \includegraphics[width= \textwidth ]{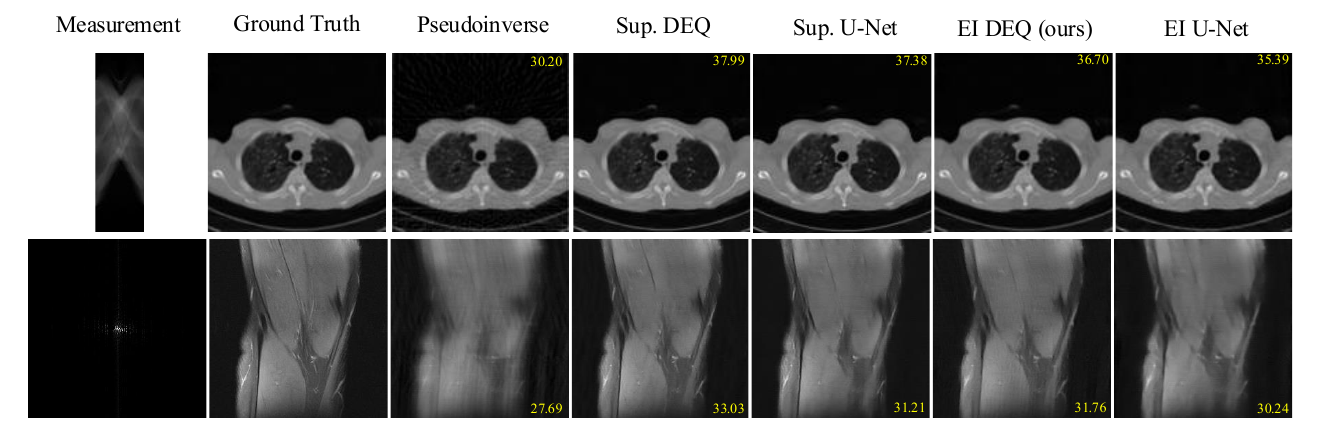}
    \vspace{-1em}
    \captionof{figure}{
    Reconstructions from Deep Equilibrium Models (DEQs) and U-Nets trained with full supervision (Sup.) and without ground truth via equivariant imaging (EI). 
    \textbf{Top --} Sparse View CT; \textbf{Bottom --} 8x Accelerated MRI with Noise. PSNR is reported in image corners.}
\end{strip}

\begin{abstract}
Equivariant imaging (EI) enables training signal reconstruction models without requiring ground truth data by leveraging signal symmetries.
Deep equilibrium models (DEQs) are a powerful class of neural networks where the output is a fixed point of a learned operator.
However, training DEQs with complex EI losses requires implicit differentiation through fixed-point computations, whose implementation can be challenging. We show that backpropagation can be implemented modularly, simplifying training. Experiments demonstrate that DEQs trained with implicit differentiation outperform those trained with Jacobian-free backpropagation and other baseline methods. Additionally, we find evidence that EI-trained DEQs approximate the proximal map of an invariant prior.
\end{abstract}
\begin{keywords}
Inverse Problems, Deep Equilibrium Models, Equivariant Imaging, Computational Imaging
\end{keywords}

\section{Introduction}
Many problems in signal processing can be formulated as inverse problems, where the goal is to recover the signal $x \in \R^n$ from the measurement $y \in \R^m$ under the forward model
\begin{equation}
    y = Ax + \epsilon, \label{eq:forwardModel}
\end{equation}
where $A: \R^n \to \R^m$ is the measurement operator and $\epsilon\in \R^m$ is noise.
A common challenge is that $A$ ill-conditioned or rank-deficient, making the inverse problem challenging due to the lack of a unique or stable solution.
Traditional methods solve inverse problems by %
minimizing the sum of a measurement-consistency term and a handcrafted prior %
using an iterative algorithm. %
Modern approaches incorporate data-driven priors, either as fixed modules within classical iterative algorithms---such as in plug-and-play (PnP) priors~\cite{venkatakrishnan2013plug,kamilovPlugandPlayPriorsApproach2017}---or as end-to-end learnable models, as in deep unrolling~\cite{gregor2010learning}.

Deep equilibrium models (DEQ) formulate reconstruction as a fixed point of a learned operator~\cite{baiDeepEquilibriumModels2019,giltonDeepEquilibriumArchitectures2021}.
Unlike deep unrolling, DEQ supports effectively infinite iterations while retaining memory efficiency and trainability via implicit differentiation~\cite{baiDeepEquilibriumModels2019}. Although PnP methods also yield fixed-point solutions~\cite{ryuPlugandPlayMethodsProvably2019}, their operators are not learned end-to-end, limiting task-specific adaptability.
DEQ overcomes this limitation and achieves state-of-the-art performance across a range of imaging inverse problems ~\cite{giltonDeepEquilibriumArchitectures2021,gungor2024deqmpi,hammernik2023physics,zhao2023deep}.

A major limitation of end-to-end models like deep unrolling and DEQ is their reliance on paired training data. In many practical settings, such as medical or scientific imaging, ground truth signals are unavailable. Equivariant Imaging (EI) enables self-supervised training using only measurements by leveraging known symmetries in the signal set (e.g., rotations, translations) to define consistency-based losses~\cite{dongdongchenEquivariantImagingLearning2021}.

Training DEQs under the EI framework can offer the best of both worlds: task-adapted models trained without requiring ground truths and the representational capacity of deep implicit networks. However, incorporating complex, transformation-based EI loss functions into DEQ training presents technical challenges due to the need for differentiating through the fixed point computations multiple times.
Although Jacobian-free backpropagation~\cite{fungJFBJacobianFreeBackpropagation2021} can avoid the complexity of implicit differentiation, it introduces approximation errors that can degrade trained model performance.

In this work, we show that a modular implementation of implicit differentiation can simplify training DEQ using complex loss functions, including those arising from EI. 
Although prior work has applied this modular implementation for supervised learning in code~\cite{giltonDeepEquilibriumArchitectures2021}, there lacks a formal explanation or application to more complex loss functions.
We hope that our formalization clarifies the use of implicit differentiation in practice and shows its useful application to self-supervised settings.
We further show that the resulting DEQ trained via EI exhibits equivariance, suggesting that it implicitly learns a proximal map corresponding to a symmetry-invariant prior function.
In summary, we make the following contributions.

\begin{itemize}[leftmargin=*]
    \item \textbf{Modular Implicit Differentiation:} We show that backpropagation for DEQ using implicit differentiation can be implemented in modular fashion, enabling training with complex loss functions involving multiple passes through the DEQ.
    \item \textbf{Implicit Differentiation vs.\ Jacobian-Free Backpropagation:} Our experimental results show that DEQs trained using implicit differentiation outperform those trained using Jacobian-free backpropagation for both supervised and EI loss functions.
    \item \textbf{Equivariant DEQ:} We show that DEQs trained with EI are equivariant, indicating that the learned operator behaves as the proximal map of an invariant prior.
\end{itemize}

\section{Background} \label{sec:background}
  
\subsection{Equivariant Imaging}
In supervised learning, we seek a reconstruction map $f_\theta$ parameterized by $\theta$ such that the reconstruction approximates the ground truth: $f_{\theta}(y) := \what{x}(\theta) \approx x$. 
We focus on learned reconstruction maps in this paper and often drop $\theta$, writing simply $\what{x}$.
Given a paired dataset $\{(x_i, y_i)\}_{i = 1}^N$, supervised training requires minimizing the objective 
\begin{equation}
    \mathcal{L}_{\text{SUP}}(\theta) = \sum_i \lsup(f_{\theta}(y_i), x_i),     \label{eq:loss}
\end{equation}
with respect to $\theta$, where $\lsup$ is a supervised loss function for an individual sample, and $\Lsup(\theta)$ is the corresponding total loss.

Given the challenge of acquiring ground truth $x$ in many settings, the EI framework proposes end-to-end training of a reconstruction map $f_{\theta}$ using only measurement data~\cite{dongdongchenEquivariantImagingLearning2021}.
Suppose the signals of interest belong to a set $\cX$ that is invariant under a group of unitary transformations $\cG$, \ie, $T_g x \in \cX$ for any $x \in \cX$ and any $g \in \cG$. 
Then, EI proposes training $f_\theta$ by minimizing the objective
\begin{equation}
\Lei(\theta) = \sum_{i = 1}^N \lmc(\what{x}_{i},\, y_i) + \alpha \sum_{i = 1}^N \sum_{g \in \cG} \lequ(T_g \what{x}_{i},\, f_{\theta}(A T_g \what{x}_{i})),
\label{eq:eiLoss}
\end{equation}
where $\what{x}_{i}(\theta) = f_\theta(y_i)$, and $\alpha > 0$ is a hyperparameter.
The loss $\lmc$ enforces measurement consistency, such as $\|Ax-y\|^2_2$, while $\lequ$ ensures equivariance of the system $f_\theta \circ A$ to the group $\cG$.
When $A$ is \emph{not} equivariant to $\cG$, applying $\{T_g\}_{g \in \cG}$ introduces virtual measurement operators $\{A T_g\}_{g \in \cG}$ to the training and enables $f_\theta$ to reconstruct signals outside the range space of $A$~\cite{dongdongchenEquivariantImagingLearning2021}. 
\subsection{Deep Equilibrium Models}

A DEQ is a neural network where the output is a fixed point of a learnable operator~\cite{baiDeepEquilibriumModels2019}.
Given a measurement $y$ from the forward model~\eqref{eq:forwardModel}, a DEQ returns a reconstruction $\what{x}(\theta)$ satisfying
\begin{equation}
    \what{x} = \F_\theta(\what{x}; y), \label{eq:fixedPoint}
\end{equation}
where $\F_\theta$ is a learnable operator parameterized by $\theta$.
In the context of inverse problems, designing $\F_\theta$ based on an iteration of a PnP method has been shown to improve the reconstruction accuracy~\cite{giltonDeepEquilibriumArchitectures2021}. For example, consider the measurement-consistency loss $\lmc(x, y)=\|Ax-y\|^2$ for the forward model~\eqref{eq:forwardModel}. Using the PnP proximal gradient method~\cite{kamilovPlugandPlayPriorsApproach2017} results in
\begin{align}
    \F_\theta(x; y) &= \D_\theta(x - \eta \nabla_x \lmc(x, y)) \label{eq:pgm} \\
    &= \D_\theta(x - \eta A^\top (Ax - y)),
\end{align}
where $\D_\theta$ is the learnable module that replaces the fixed denoiser in PnP PGM, and $\eta > 0$ is a step size.
The fixed point of $\F_\theta$ can be computed through fixed-point iterations or using acceleration techniques~\cite{baiDeepEquilibriumModels2019,giltonDeepEquilibriumArchitectures2021} such as Broyden's method~\cite{broyden1965class} or Anderson acceleration~\cite{walkerAndersonAccelerationFixedPoint2011}.

For a DEQ, $f_\theta(y) = \Fix_{\what{x}} \{\F_\theta(\what{x}; y)\}$,
where $\Fixeq{x}{f(x)}$ denotes the fixed point $x^*$ satisfying $x^* = f(x^*)$.
Training a DEQ in supervised fashion by minimizing~\eqref{eq:loss} with a gradient based method is challenging, because backpropagation through fixed-point iterations typically incurs an intractably large memory footprint. 
To address this, we can exploit the fixed-point condition~\eqref{eq:fixedPoint} to rewrite the gradient of a loss $\ell(\what{x})$ with respect to $\theta$ as
\begin{align}
    \pfrac{\ell(\what{x})}{\theta} = \pfrac{\ell(\what{x})}{\what{x}} \left(I - \pfrac{\F_\theta(x)}{x} \Big\vert_{x = \what{x}} \right)^{-1} \pfrac{\F_\theta(\what{x})}{\theta}, \label{eq:dldt}
\end{align}
where, for brevity, we drop $y$ in $\F_\theta(x; y)$~\cite{baiDeepEquilibriumModels2019,giltonDeepEquilibriumArchitectures2021}.
Explicitly computing the inverse matrix in~\eqref{eq:dldt} is generally intractable for high-dimensional signals.
Implicit differentiation (ID) and Jacobian-free backpropagation (JFB) are two main approaches to overcome this problem.
In ID, the gradient computation is reformulated as solving an auxiliary fixed-point problem:
\begin{equation}
    \pfrac{\ell(\what{x})}{\theta} = \Fixeq{\beta}{\beta \pfrac{\Ftx}{x} \Big \vert_{x = \what{x}} + \pfrac{\ell(\what{x})}{\what{x}}} \pfrac{\F_\theta(\what{x})}{\theta}, \label{eq:id}
\end{equation}
which can be solved using fixed-point iteration methods such as Broyden's method~\cite{broyden1965class} and Anderson acceleration~\cite{walkerAndersonAccelerationFixedPoint2011}. 

With JFB, the inverse matrix term is entirely dropped and the resulting vector 
$\widetilde{\pfrac{\ell}{\theta}} := \pfrac{\ell}{\what{x}} \pfrac{\F_\theta(\what{x})}{\theta}$,
which is a descent direction for $\theta$ under certain conditions,
is used for backpropagation instead of the true gradient~\cite{fungJFBJacobianFreeBackpropagation2021}.
However, as shown later in \Cref{tab:benchmark}, we find that DEQs trained using JFB achieve less accurate reconstructions compared to ID-trained DEQs.

\section{Backpropagation for DEQ}

\subsection{Difficulty in Training DEQ with EI}

Training a DEQ $f_\theta(y) = \Fix_{\what{x}} \{\F_\theta(\what{x}; y)\}$ by minimizing the EI objective~\eqref{eq:eiLoss} is complicated by DEQ's fixed point computation and multiple passes through $f_\theta(y)$.
Consider the equivariant loss $\lequ(x_2, x_3)$, where $x_2 = T_g f_\theta(y)$ and $x_3 = f_{\theta}(A x_2)$.
The gradient of $\lequ(x_2, x_3)$ with respect to $\theta$ is
\begin{equation}
    \pfrac{\lequ}{\theta} = \pfrac{\lequ}{x_2} \pfrac{x_2}{\theta} + \pfrac{\lequ}{x_3} \left(\pfrac{x_3}{\theta} + \pfrac{x_3}{x_2} \pfrac{x_2}{\theta} \right).
\end{equation}
Each of the three terms involves a derivative of a DEQ fixed point with respect to $\theta$.
In principle, we could apply ID separately to each term by deriving the corresponding fixed-point equation, similarly to equation~\eqref{eq:id}.
However, as detailed in \Cref{sec:id-deq}, the gradient computation for all terms can be implemented modularly, using the same core code, without needing to manually formulate and solve three distinct fixed-point subproblems.

\subsection{Modular Implicit Differentiation} \label{sec:id-deq}

We can simplify ID for complex loss functions such as~\eqref{eq:eiLoss} by leveraging the fact that the vector-Jacobian product between an upstream gradient and the Jacobian of a DEQ's fixed point with respect to its parameters can be systematically reduced to a fixed-point equation. This result is formalized in the following proposition.
\begin{proposition} \label{prop:modularID}
The vector-Jacobian product between a row vector $g \in \R^{1 \times n}$ and a Jacobian $\pFrac{\what{x}}{\theta} \in \R^{n \times p}$ of a fixed point $\what{x} = \Ftwx$ with respect to the parameters $\theta \in \R^n$ is
\begin{equation}
    g \pfrac{\what{x}}{\theta} = \Fixeq{\beta}{\beta \pfrac{\Ftx}{x} \Big\vert_{x = \what{x}} + g} \pfrac{\Ftwx}{\theta}. \label{eq:modularID}
\end{equation}
\end{proposition}
We represent gradients as row vectors to avoid unnecessary transposes, simplifying notation. 
The standard ID formulation~\eqref{eq:id} is a direct consequence of \Cref{prop:modularID} when $g = \partial \ell / \partial \what{x}$.
This proposition enables a modular implementation of ID: any vector–Jacobian product involving the DEQ’s output can be replaced with the fixed-point computation in~\eqref{eq:modularID}, followed by a multiplication with $\partial \Ftwx / \partial \theta$. The PyTorch implementation shown in \Cref{fig:id-code} directly realizes this computation.

Although this modular implementation of ID was previously used in code for supervised reconstruction with DEQs~\cite{giltonDeepEquilibriumArchitectures2021}, to the best of our knowledge, \Cref{prop:modularID} provides the first mathematical formalization of this technique, and our work is the first to apply it to the self-supervised EI setting. 
\footnotetext[1]{Adapted from the code in \href{https://github.com/dgilton/deep_equilibrium_inverse/blob/main/solvers/new_equilibrium_utils.py\#L242}{Gilton et al.}}

\begin{figure}[t]
\centering
\begin{lstlisting}[
language=Python,
basicstyle=\ttfamily\footnotesize,
numbers=left,
numbersep=5pt,
backgroundcolor=\color{LightGray},
frame=lines,
rulecolor=\color{black},
xleftmargin=2em,
framexleftmargin=1.5em,
breaklines=true
]
# F(x, y) is the learnable operator.
# solver(func) yields a fixed point x=func(x).
def forward(y):
    with torch.no_grad():
        xhat = solver(lambda x: F(x, y))
    xhat = F(xhat, y)
    x2 = xhat.clone().detach().requires_grad_()
    Fx = F(x2, y)
    def backward_hook(grad):
        bstar = solver(lambda b: 
            autograd.grad(Fx, x2, b)[0] + grad)
        return bstar
    xhat.register_hook(backward_hook)
    return xhat
\end{lstlisting}
\caption[Caption]{\textbf{PyTorch implementation of DEQ forward pass and implicit differentiation}, adapted from~\cite{giltonDeepEquilibriumArchitectures2021}\protect\footnotemark.
The fixed-point is computed in lines 3–4; the remaining lines implement ID.
Lines 7-12 define \texttt{backward\_hook} for backpropagating through the fixed point \texttt{xhat} given an upstream gradient \texttt{grad}.
Line 6 performs an evaluation of $\F_\theta$ so that $\partial \Ftwx / \partial \theta$ is captured in the autograd graph.
\vspace{-1.5em}
}
\label{fig:id-code}
\end{figure}

\section{Experiments}

We evaluate the performance of DEQs trained using the EI framework with ID against a range of reconstruction methods. These include DEQs trained with supervision, DEQs trained using JFB, as well as several baseline methods: deep unrolled models, U-Net refinement, PnP priors, and Regularization by Denoising (RED)~\cite{romano2017little}.

\subsection{Tasks}

We consider two representative inverse problems: sparse-view computed tomography (CT) and accelerated magnetic resonance imaging (MRI). Both imaging systems follow the linear forward model in~\eqref{eq:forwardModel}. 
For EI training, we define the transformation group $\mathcal{G}$ as rotations in 1-degree increments ($|\mathcal{G}| = 360$) for both tasks. Since the forward operators for accelerated MRI and sparse-view CT are not equivariant to rotations, each $AT_g$ has a different null-space, enabling learning without access to any ground truth. 
Each dataset is split into 90\% for training and 10\% for testing.
We use the \textit{Deep Inverse} library~\cite{tachella2025deepinverse} for the implementation of measurement operators $A$ and related operators such as pseudoinverses $A^\dagger$. We use \textit{TorchDEQ} for the implementation of DEQs~\cite{gengTorchDEQLibraryDeep2023}. 

\subsubsection{Sparse-view CT}

The measurement operator $A$ is the discrete Radon transform, subsampled over 50 projection angles uniformly spaced between \SI{0}{\degree} and \SI{180}{\degree}. We use the CT100 dataset~\cite{CT_Dataset}, which consists of 100 CT images, each resized to size $128 \times 128$. The forward model is assumed to be noiseless ($\epsilon = 0$). We found equivariance strength of $\alpha = 100$ was best for all models based on hyperparameter search. 

\subsubsection{Accelerated MRI}

The task is $8\times$ accelerated single-coil MRI. The measurement operator $A$ is a subsampled 2D discrete Fourier transform with a fixed Gaussian mask. The mask fully samples low-frequency components and randomly samples high-frequency components based on a Gaussian distribution~\cite{schlemper2017deep}. 
The real and complex parts of the measurements are fed to two channels for all models.
Additive white Gaussian noise with standard deviation $0.05$ is added to the measurements. %
The dataset consists of 199 knee scan slices with size $320 \times 320$ from the fastMRI dataset~\cite{fastMRIZbontar2018Nov}. We used $\alpha = 1$ for all MRI experiments based on hyperparameter search. Similar to CT, we found that the best $\alpha$ value didn't vary across U-Nets or DEQs. 

\subsection{Reconstruction Methods}
DEQs, deep unrolling models, and the U-Net models are trained end-to-end, while PnP prior and RED use a pre-trained denoiser without end-to-end training.
\begin{table*}[ht]
    \centering 
    \fontsize{9pt}{11pt}\selectfont
    \resizebox{\textwidth}{!}{
    \begin{tabular}{l l *{9}{c}}
        \toprule
         Task & Method & \multicolumn{2}{c}{DEQ} & \multicolumn{1}{c}{U-Net} & \multicolumn{4}{c}{Unrolled} & PnP/RED \\
         \cmidrule(r){3-4} \cmidrule(r){5-5} \cmidrule(r){6-9} \cmidrule(r){10-10}
         & & ID & JFB & \(A^{\dagger}y\) & 1-Layer & 3-Layer & 5-Layer & 7-Layer & DnCNN \\
        \midrule
        \textbf{\multirow{2}{*}{\begin{tabular}{@{}c@{}}MRI\end{tabular}}} & Supervised & \textbf{32.61} & \underline{32.51} & 30.587 & 29.16  & 29.45 & 29.33 & 29.57 & \multirow{2}{*}{25.07} \\
         & Self-Supervised & \textbf{31.34} & \underline{29.39}& 28.02 & 27.35 & 27.56 & 27.12 & 27.24 & \\
        \midrule
        \textbf{\multirow{2}{*}{\begin{tabular}{@{}c@{}}CT\end{tabular}}} & Supervised & \textbf{36.77} & 32.80 & \underline{36.33} & 35.14 & 34.57&  34.99&  35.80& \multirow{2}{*}{31.90} \\
         & Self-Supervised & \textbf{35.41} &33.52 &  34.03 & 31.73 & 32.01& 32.34& \underline{34.66} & & \\
        \bottomrule
    \end{tabular}
    }
    \caption{\textbf{Test PSNR.} 
    The best result is in bold and the second best is underlined. DEQ trained with ID outperforms U-Net and DEQ with JFB. Despite having access to no ground truth data, EI performance is close to supervised. }
    \label{tab:benchmark}
\end{table*}

\vspace{-0.5em}
\subsubsection{Deep Equilibrium Models}

We use different iterative reconstruction methods as templates for the two tasks to demonstrate wide applicability of modular ID.
For CT, we use RED as the template, resulting in the DE-Grad architecture~\cite{giltonDeepEquilibriumArchitectures2021}:
\begin{equation}
    \F_\theta(x) = x - \eta \nabla \lmc(x) - \eta (x - \operatorname{D}_\theta(x)).
\end{equation}
    
For MRI, we use PGM as the template, resulting the fixed-point operator defined in~\eqref{eq:pgm}, \ie, the DE-Prox architecture~\cite{giltonDeepEquilibriumArchitectures2021}.
For both DEQs, the learnable module $\operatorname{D}_\theta$ is a residual U-Net denoiser~\cite{ronneberger2015u}, whose initial weights are pretrained on the BSD500 dataset~\cite{MartinFTM01}.
To ensure that $\F_\theta$ is contractive and admits a fixed point, we follow standard practice by applying spectral normalization or weight normalization~\cite{salimans2016weight}
and using group normalization layers~\cite{wu2018group}.

\vspace{-0.5em}
\subsubsection{Deep Unrolling}

We implement deep unrolled models based on the iterative algorithm corresponding to each task: RED for CT and PGM for MRI; unrolled PGM is also known as LISTA~\cite{gregor2010learning}. 
Model weights in all iterations are shared to keep the number of parameters consistent with DEQs.
The deep unrolled models use the same initial U-Net denoiser as our DEQ experiments.

\vspace{-0.5em}
\subsubsection{U-Net Refinement}

We include a simple U-Net baseline reconstruction that refines the pseudoinverse solution, \ie, $f_\theta(y) = \operatorname{D}_\theta(A^\dagger y)$, following experiments in~\cite{dongdongchenEquivariantImagingLearning2021}.

\vspace{-0.5em}
\subsubsection{PnP / RED}

These classical iterative methods use pretrained denoisers without end-to-end training. We use RED~\cite{romano2017little} for CT and PnP PGM~\cite{kamilovPlugandPlayPriorsApproach2017} for MRI. Both use a DnCNN denoiser~\cite{zhang2017beyond} with weights from the Deep Inverse library~\cite{tachella2025deepinverse}.
We use DnCNN instead of U-Net as we found that U-Net performs poorly as a denoiser and yields worse reconstruction when used in PnP and RED.

\subsection{Results}

\subsubsection{ID yields better models than JFB}
As shown in \Cref{tab:benchmark}, DEQs trained with ID consistently outperform those trained with JFB. Although JFB offers computational convenience, the lack of exact gradient information leads to suboptimal convergence and inferior reconstruction quality.

\subsubsection{EI performance is close to supervised}
Using ID, DEQs trained with EI attains a PSNR within 1.3 dB of that achieved by supervised training.
This result demonstrates the effectiveness of EI as a self-supervised approach for training powerful reconstruction models when ground truth is unavailable.
Additionally, both DEQ and U-Net models trained using EI have higher test PSNR than denoiser-based methods like PnP and RED.

\subsubsection{DEQ has the strongest performance}
DEQ-based methods achieve the strongest overall performance across both tasks, outperforming unrolled networks, U-Net refinement, and PnP/RED.
These results highlight the strength of DEQ in capturing the structure of inverse problems more effectively than deep unrolling or PnP/RED.

\section{Discussion}

\subsection{Imaging Pipeline is Equivariant}

\begin{table}[h] 
\centering
\begin{tabular}{l|c|ccc}
\hline
& Method & \textbf{Equivariant Imaging} & \textbf{Supervised}\\
\hline
MRI & DEQ & $34.64 \pm 2.449$ & $34.66 \pm 2.340$ &\\
\hline
MRI & U-Net & $ 28.01 \pm 1.311$ & $25.62 \pm 1.244 $ &\\
\hline
CT  & DEQ & $47.57 \pm1.268$ & $39.24\pm 2.734$ & \\
\hline
CT  & U-Net & $42.55 \pm 1.449 $& $36.48 \pm 1.989$ & \\
\hline
\end{tabular}
\caption{\textbf{Equivariance error of imaging pipeline} $f_\theta \circ A$, quantified by PSNR of the difference between $f_\theta(A T_g x)$ and $T_g f_\theta(A x)$. }
\label{tab:systemEquivariance}
\end{table}

If the signal set is invariant under a group of transformations $\cG$, then it is natural to expect the entire measurement-reconstruction system $f_\theta \circ A$ to be equivariant with respect to $\cG$~\cite{dongdongchenEquivariantImagingLearning2021}. For instance, a reconstruction from an MRI scan of a rotated patient should correspond to the rotation of the original reconstruction.

Results in \Cref{tab:systemEquivariance} demonstrates this system equivariance property.
We quantify the equivariance of $f_\theta \circ A$ by computing the average difference between $f_\theta(A T_g x)$ and $T_g f_\theta(A x)$ across test set images $x$, reporting the result in terms of PSNR (referred to as \emph{equivariance error}). While all ngles are used during training, we only choose $T_g$ to be a \SI{90}{\degree} counterclockwise rotation to avoid aliasing edge artifacts.

EI-trained models have significantly higher equivariance error than those from supervised training. We attribute this observation to the training data, which are not augmented by rotated images, whereas EI implicitly augments rotated images during training.

\subsection{EI-trained DEQ Operator is Equivariant}

\begin{table}[ht] %
\centering
\begin{tabular}{l|ccc}
\hline
 & \textbf{EI} & \textbf{Supervised} & \textbf{Initialization}\\
\hline

MRI & $39.96 \pm 2.333$ & $32.59 \pm 2.062$ & $ 37.50 \pm 0.564$\\
\hline

CT  & $52.44 \pm 1.412 $& $46.75 \pm 2.056$ & $33.94 \pm 0.768$\\
\hline
\end{tabular}
\caption{\textbf{Equivariance error of DEQ operator $\F_\theta$}, quantified by PSNR of the difference between $T_g \F_\theta(x, Ax)$ and $\F_\theta(T_g x, A T_g x)$. Initialization refers to a DEQ where $\operatorname{D}_\theta$ is the pretrained denoiser used at the start of training.}
\label{tab:deqEquivariance}
\end{table}

Training DEQs using the EI framework also reveals a deeper structural property: the learned operator $\F_\theta$ becomes equivariant to the transformation group $\cG$.
As shown in \Cref{tab:deqEquivariance}, DEQs trained with EI exhibit significantly stronger equivariance than those trained with standard supervised learning. This is quantified by the PSNR of the difference between $T_g \F_\theta(x, Ax)$ and $\F_\theta(T_g x, A T_g x)$ averaged over test images. Again, we evaluate equivariance $T_g$ with a \SI{90}{\degree} rotation to avoid edge artifacts.

This result suggests that the DEQ's learnable module $\D_\theta$ approximates the proximal map of a $\cG$-invariant prior function, as the proximal map of an invariant function is equivariant under unitary group actions~\cite{celledoni2021equivariant}.
The equivariant loss $\lequ$ in the EI objective~\eqref{eq:eiLoss} is invariant to $\cG$ when averaged over group elements $g \in \cG$, further supporting the interpretation that the DEQ implicitly learns an invariant prior.

This observation highlights a deeper connection between explicit regularization and learned reconstruction maps, which is a major theoretical question. 
Prior works have explored this relationship in various contexts~\cite{romano2017little,zouDeepEquilibriumLearning2023,hurault2021gradient,goujon2023neural,fang2023s}.

\section{Conclusion}

We formalize a modular implementation of implicit differentiation for training DEQs and demonstrate the performance gain of training DEQs with implicit differentiation when compared to Jacobian-free backpropagation and other reconstruction methods.
We also show that the resulting EI-trained operator in DEQ is equivariant, suggesting that DEQ implicitly learns an invariant prior function.

Given the superior reconstruction quality of DEQs, developing more robust and efficient methods for training DEQs, especially in self-supervised settings, remains a critical direction for future work, as it unlocks the potential to use DEQs in domains where ground truth data are limited or unavailable.
Although our results suggest that DEQ implicitly learns an invariant prior function, our evidence is not conclusive. A rigorous characterization of the implicit prior is another future direction.

\bibliographystyle{IEEEbib}
\bibliography{abbrv,final_ref}

\end{document}